\documentclass[11pt]{article}

\usepackage[margin=1in]{geometry}
\usepackage{amsmath,amssymb,amsfonts,amsthm,bm,mathtools}
\usepackage{graphicx}
\usepackage{booktabs}
\usepackage{xcolor}
\usepackage[colorlinks=true,linkcolor=blue,citecolor=blue,urlcolor=blue]{hyperref}
\hypersetup{pdftitle={Krylov Break Times from an Inhomogeneous Lieb--Robinson Light Cone},pdfauthor={S. Matsuura, Y. Kawamura, J. Salfi, S. Iso}}
\usepackage{microtype}
\usepackage{authblk}
\usepackage{enumitem}
\usepackage{tikz}
\setlist{nosep}

\theoremstyle{plain}
\newtheorem{theorem}{Theorem}
\newtheorem{proposition}{Proposition}

\theoremstyle{definition}
\newtheorem{definition}{Definition}

\newcommand{\HH}{\mathcal{H}}
\newcommand{\KK}{\mathcal{K}}

\newcommand{\R}{\mathbb{R}}

\newcommand{\ii}{\mathrm{i}}
\newcommand{\dd}{\mathrm{d}}
\newcommand{\e}{\mathrm{e}}

\newcommand{\vlr}{v_{\mathrm{LR}}}
\DeclareMathOperator{\arsinh}{arsinh}

\title{Krylov Break Times \\ from an Inhomogeneous Lieb--Robinson Light Cone}

\author[1,2,3,4]{Shunji Matsuura}
\author[4]{Yoji Kawamura}
\author[2,5,6]{Joseph Salfi}
\author[1,7,8]{Satoshi Iso}

\affil[1]{RIKEN Center for Interdisciplinary Theoretical and Mathematical Sciences (iTHEMS), RIKEN, Wako, Saitama 351-0198, Japan}
\affil[2]{Department of Electrical and Computer Engineering, University of British Columbia, Vancouver, BC V6T 1Z4, Canada}
\affil[3]{Department of Physics, University of Guelph, Guelph, ON N1G 2W1, Canada}
\affil[4]{Center for Mathematical Science and Advanced Technology, Japan Agency for Marine-Earth Science and Technology, Yokohama 236-0001, Japan}
\affil[5]{Department of Physics and Astronomy, University of British Columbia, Vancouver, BC V6T 1Z4, Canada}
\affil[6]{Stewart Blusson Quantum Matter Institute, University of British Columbia, Vancouver, BC V6T 1Z4, Canada}
\affil[7]{KEK Theory Center, Institute of Particle and Nuclear Studies, Oho 1-1, Tsukuba, Ibaraki 305-0801, Japan}
\affil[8]{Graduate University for Advanced Studies (SOKENDAI), Oho 1-1, Tsukuba, Ibaraki 305-0801, Japan}

\date{\today}

\begin{document}

\maketitle

\begin{abstract}
Krylov and Lanczos approximations are used in quantum dynamics, quantum subspace methods, and Hamiltonian learning. A practical question is how long an $m$-dimensional Krylov truncation can be trusted. We argue that this time is fixed by causal propagation on the associated Jacobi chain. The relevant distance is not the Krylov index itself, but the inhomogeneous transport metric
\[
   \rho(m,n) = \sum_{j=\min(m,n)}^{\max(m,n)-1} \frac{1}{b_j},
\]
where $b_j$ is the Lanczos hopping across the bond $j \leftrightarrow j+1$. We prove a Lieb--Robinson bound in this metric. Its small-weight limit gives the velocity $\vlr = 2$, meaning that propagation is exponentially suppressed outside the cone $\rho(m,n) \simeq 2|t|$. The error of a finite Krylov approximation to the return amplitude is a round-trip effect: information has to travel from the probe to the truncation boundary and back. 
Combining the Lieb--Robinson bound with Duhamel's formula yields a lower bound on the error of the truncated dynamics. For a fixed tolerance $\epsilon$, let the \emph{break time} $t_\ast(m;\epsilon)$ denote the longest time for which the $m$-dimensional truncation is guaranteed to reproduce the exact return amplitude within error $\epsilon$. We show that
\[
   t_\ast(m;\epsilon) \ge \tau_m[1-o(1)], \qquad
   \tau_m = \rho(0,m) = \sum_{j<m}\frac{1}{b_j}.
\]
The factor of two in the round-trip distance cancels the Lieb--Robinson velocity. When the probe spreads along the chain, this lower bound is also tight, so $t_\ast(m) \simeq \tau_m$. 
The situation is different when the probe excites only a localized part of the spectrum, or a part already resolved by the truncation. In this case, essentially no signal reaches the boundary. Beyond a state-dependent Krylov dimension $m_\ast$, the approximation can therefore remain accurate at all times, and the break time is effectively infinite.
Numerical tests on spin chains and random Jacobi matrices support $t_\ast(m) \simeq \tau_m$ in the transport-limited regime.
\end{abstract}

\section{Introduction} \label{sec:intro}

Lanczos and Krylov methods are often used because they replace a large Hamiltonian problem with a much smaller tridiagonal one. Starting from a state $|v_0\rangle$, the Lanczos recursion constructs the part of the Hamiltonian that is visible from this state.
The resulting Jacobi matrix is not merely a numerical device. It is also the exact generator of the survival amplitude within the cyclic subspace generated by $|v_0\rangle$. This cyclic subspace is precisely the Krylov subspace
\[
\KK(H,v_0)
=
\mathrm{span}\{|v_0\rangle, H|v_0\rangle, H^2|v_0\rangle, \dots\}.
\]
Throughout this paper, we use the terms cyclic subspace and Krylov subspace interchangeably. This identification is made precise in Section~\ref{sec:jacobi}.
This naturally raises a basic question. If we stop the recursion at dimension $m$, for how long is the resulting dynamics reliable?

This question appears in several forms. In classical numerical analysis it is the time window of a Krylov approximation to the matrix exponential. In quantum algorithms it is the depth or subspace dimension needed to reproduce a target time evolution. In spectroscopy and restricted Hamiltonian learning~\cite{Burgarth2009b,DiFranco2009,Zhang2014}, it is the question of how much of the measured signal can be predicted from a finite learned Jacobi model. In all cases, one would like a stopping criterion that is available while the Lanczos recursion is being performed. Such a criterion should use the already computed Lanczos coefficients, not the full spectrum of the original Hamiltonian.

Let $H$ be a time-independent Hamiltonian and $|v_0\rangle$ a normalized probe state. We consider the survival amplitude
\begin{equation}
   A(t) = \langle v_0 | \e^{-\ii Ht} | v_0 \rangle.
\end{equation}
Lanczos recursion maps this amplitude to the return amplitude on a Jacobi chain,
\begin{equation}
   A(t) = \langle 0 | \e^{-\ii Jt} | 0 \rangle,
\end{equation}
where $J$ is the Jacobi matrix of the pair $(H,|v_0\rangle)$ and $|0\rangle$ is its origin (chain-site) vector; the chain-site basis $\{|n\rangle\}$, with $|0\rangle = |v_0\rangle$, is defined in Section~\ref{sec:jacobi}. The $m$-dimensional Krylov approximation replaces $J$ by the principal truncation $J_m$ and uses
\begin{equation}
   A_m(t) = \langle 0 | \e^{-\ii J_m t} | 0 \rangle.
\end{equation}
We measure its validity by the following break time.

\begin{definition}[Break time] \label{def:breaktime}
For a tolerance $\epsilon > 0$, define
\begin{equation}
   t_\ast(m;\epsilon) = \inf \Big\{ t \ge 0 \colon \max_{0 \le s \le t}| A(s) - A_m(s) | > \epsilon \Big\}. \label{eq:breaktime}
\end{equation}
\end{definition}

\noindent
The prefix maximum is used to make $t_\ast$ the first time up to which the whole interval $[0,t]$ is reliable. Thus $t < t_\ast$ means that the approximation has stayed within the tolerance at all earlier times.

The main result is that this break time is controlled by the transport time
\begin{equation}
   \tau_m = \sum_{j=0}^{m-1} \frac{1}{b_j},
\end{equation}
where $b_j$ is the hopping amplitude on the Jacobi-chain bond $j \leftrightarrow j+1$. More precisely, under boundedness assumptions stated below,
\begin{equation}
   t_\ast(m;\epsilon) \ge \tau_m [1 - o(1)]. \label{eq:mainbound}
\end{equation}
By Definition~\ref{def:breaktime}, $t_\ast(m;\epsilon)$ is exactly the time at which the truncation breaks down: for every $t < t_\ast(m;\epsilon)$ the $m$-dimensional approximation reproduces $A(t)$ to within $\epsilon$, whereas for $t > t_\ast(m;\epsilon)$ the error has already exceeded $\epsilon$ and the approximation can no longer be trusted. Equation~\eqref{eq:mainbound} therefore guarantees that this reliable window is at least $\tau_m$ long. For ballistic Jacobi chains~\footnote{We call a Jacobi chain \emph{ballistic} when an excitation travels along it at a finite, nonzero speed, so that the distance it covers grows linearly in time. In the transport metric this is the case whenever the hoppings $b_j$ stay bounded away from zero, so that $\tau_m$ grows linearly with $m$; the free chain $b_j = b_\infty$ is the simplest example.} this lower bound is asymptotically sharp. In this regime a further Lanczos step increases the trusted time window by about $1/b_m$. Whether this lower bound is also an \emph{upper} bound, in which case $t_\ast(m) \simeq \tau_m$, is a separate question. It holds only when the relevant spectral component actually transports, which needs spectral information that the Lieb--Robinson bound alone does not provide. We return to this and to the opposite regime of spectral capture, in Section~\ref{sec:saturation} and Appendix~\ref{app:refined}.

The transport mechanism is controlled by a light-cone structure in Krylov space. The Jacobi chain is one-dimensional with only nearest-neighbor couplings, although these couplings are generally nonuniform. In many-body systems, the Lanczos coefficients can vary substantially as a function of the Krylov index. Consequently, the natural notion of distance is not the ordinary lattice separation $|m-n|$. We instead introduce an inhomogeneous transport metric $\rho(m,n)$, defined in Eq.~\eqref{eq:rho2} below by summing the local traversal times $1/b_j$ over the bonds connecting sites $m$ and $n$. Bonds with larger hopping amplitudes correspond to shorter distances in this metric, whereas weaker bonds correspond to longer ones. In this sense, $\rho$ measures the total local propagation time accumulated along the Jacobi chain.

We prove a Lieb--Robinson bound in this metric with velocity $2$. A finite truncation affects the return amplitude only after propagation from the initial site to the boundary and back. The relevant distance is therefore roughly $2\tau_m$. Dividing by the velocity $2$ gives the break-time scale $\tau_m$. This cancellation explains why the one-way transport time appears in a return-amplitude problem.

\paragraph{Relation to previous work.}
The Lanczos recursion and Jacobi representation are classical tools in the recursion method and in Krylov approximations to the matrix exponential~\cite{Lanczos1950,Haydock1972,Viswanath1994,Hochbruck1997}. They are also used in quantum subspace and Krylov algorithms~\cite{Klymko2022,Yoshioka2022,CortesGray2022,Shen2023,Kirby2024}. These works give the basic algorithmic and spectral framework, but they do not give a coefficient-resolved prediction of the finite-Krylov break time. Existing a priori error bounds are rigorous, but in practice they may be too pessimistic as stopping rules because they do not use the actual causal propagation on the Jacobi chain produced by the Lanczos coefficients.

The closest dynamical precursor is the work of Ruffinelli, Fortes, Larocca, and Wisniacki~\cite{Ruffinelli2022}. They showed that the Krylov truncation error can be expressed as a Loschmidt echo on the Jacobi tight-binding chain. In their picture the error starts to grow when a wave packet reaches the truncation boundary. They also introduced characteristic times such as $t_{\mathrm{exp}}$ and $t_{\mathrm{col}}$ and observed quasi-linear scaling with Krylov dimension. The present work keeps this physical picture and adds the metric and the corresponding bound. The scaling variable is not the bare dimension $m$ but the transport distance $\tau_m = \sum_{j<m} 1/b_j$. We also prove a Lieb--Robinson bound in this nonuniform metric and use a Duhamel round-trip argument to obtain~\eqref{eq:mainbound}.

Another related direction is operator Krylov complexity, in which the Lanczos recursion is applied to the Heisenberg evolution of an operator~\cite{Parker2019,Rabinovici2021,Rabinovici2025review,Nandy2025review}. For example, Alishahiha and Banerjee derived a geometric speed limit for operator Krylov amplitudes~\cite{Alishahiha2025}. That problem is not the same as the finite-dimensional state-Krylov break time defined in~\eqref{eq:breaktime}; it concerns a different Krylov space and does not directly treat the return-amplitude error caused by truncating a state-Lanczos chain. Also, a uniform metric cannot describe cases where the Lanczos hoppings vary along the chain. In the homogeneous Jacobi chain, the exact Bessel kernel already has a sharp light cone with velocity $2b$~\cite{Watson1944}. What is needed here is the inhomogeneous version and its consequence for a finite Krylov approximation.

The paper is organized as follows. Section~\ref{sec:jacobi} introduces the Jacobi chain and the transport metric. Section~\ref{sec:LR} proves the Krylov Lieb--Robinson bound and explains how the velocity $2$ is read from the small-$\lambda$ limit. Section~\ref{sec:roundtrip} applies Duhamel's formula to the return amplitude and derives the break-time lower bound. Section~\ref{sec:saturation} separates the transport-limited regime from spectral capture. Section~\ref{sec:numerics} summarizes the numerical evidence. Section~\ref{sec:discussion} discusses the stopping rule, the scope of the result, and the connection with restricted-access learning. Section~\ref{sec:conclusion} concludes the paper. Some details are placed in the appendices. Throughout the paper, we use standard asymptotic notation, $o(\cdot)$, $O(\cdot)$, and $\Theta(\cdot)$.~\footnote{In the relevant asymptotic limit, $f = o(g)$ means $f/g \to 0$; $f = O(g)$ means that $|f/g|$ remains bounded; and $f = \Theta(g)$ means that $f/g$ is bounded both above and below by positive constants.}

\section{Jacobi chain and transport time} \label{sec:jacobi}

This section fixes notation and explains the metric used later. The only point is that the Krylov index becomes a one-dimensional coordinate after Lanczos recursion, while the hopping coefficients determine the local speed along this coordinate.

Let $H$ be self-adjoint on a finite-dimensional Hilbert space, and let $|v_0\rangle$ be normalized. The cyclic subspace generated by this pair is
\begin{equation}
   \KK(H,v_0) = \mathrm{span}\{ |v_0\rangle, H|v_0\rangle, H^2|v_0\rangle, \dots \}.
\end{equation}
The Lanczos recursion orthonormalizes this sequence into the chain-site basis $\{|n\rangle\}_{n \ge 0}$ of $\KK(H,v_0)$, with $|0\rangle = |v_0\rangle$ and $\langle m|n\rangle = \delta_{mn}$. In this basis $H$ acts tridiagonally,
\begin{equation}
   H|n\rangle = \beta_n|n-1\rangle + \alpha_n|n\rangle + \beta_{n+1}|n+1\rangle,\qquad |0\rangle = |v_0\rangle,\quad \beta_0 \equiv 0.
\end{equation}
Here $\alpha_n \in \R$ and $\beta_{n+1} > 0$ until the recursion terminates. These orthonormal vectors $|n\rangle$ are the coordinate (site) vectors of the Jacobi chain; they are the only basis states used in the rest of the paper, and in Sections~\ref{sec:LR}--\ref{sec:roundtrip} the shorthand $|m\rangle,|0\rangle,|M\rangle,\dots$ always refers to them. To avoid an index ambiguity in the metric, we denote the hopping on the edge $j \leftrightarrow j+1$ by
\begin{equation}
   b_j \coloneqq \beta_{j+1} > 0.
\end{equation}
Then the restriction of $H$ to $\KK(H,v_0)$ is represented by the Jacobi matrix
\begin{equation}
   J = \begin{pmatrix}
   \alpha_0 & b_0      &          &        \\
   b_0      & \alpha_1 & b_1      &        \\
            & b_1      & \alpha_2 & \ddots \\
            &          & \ddots   & \ddots 
   \end{pmatrix}.
\end{equation}
The Jacobi matrix reproduces every probe expectation value exactly: for any function $f$ of the Hamiltonian, such as the propagator $\e^{-\ii Ht}$ (whose expectation is $A(t)$), a resolvent, or a spectral projector,
\begin{equation}
   \langle v_0|f(H)|v_0\rangle = \langle 0|f(J)|0\rangle.
\end{equation}
The finite Krylov approximation replaces $J$ by its $m \times m$ principal truncation $J_m$. We use a single truncated chain, with two index conventions for it. $J_m$ is the principal $m \times m$ truncation (sites $0,\dots,m-1$); this is the $m$-dimensional model whose break time we study, so $m$ is the number of Lanczos steps taken. In the causal arguments of Sections~\ref{sec:LR}--\ref{sec:roundtrip} it is convenient to label the same truncation by its last retained site $M$ and write $J_M$ (sites $0,\dots,M$), because the light-cone cut is placed at a definite bond; the two are the same matrix at $m = M+1$, and the one-step difference $\tau_m - \tau_M = 1/b_M$ is sublinear and absorbed into the corrections below. The unsubscripted $J$ denotes the full half-line operator.

Since $b_j$ is the hopping across the bond $j \leftrightarrow j+1$, the natural local crossing time of this bond is of order $1/b_j$. We therefore define 
\begin{equation}
   \rho(m,n) = \sum_{j=\min(m,n)}^{\max(m,n)-1} \frac{1}{b_j},\qquad \rho(n,n) = 0, \label{eq:rho2}
\end{equation}
and
\begin{equation}
   \tau_m = \rho(0,m) = \sum_{j=0}^{m-1} \frac{1}{b_j}.
\end{equation}
For a homogeneous chain $b_j = b$, this gives $\rho(m,n) = |m-n|/b$. In that case the usual tight-binding dispersion has maximal group velocity $2b$, so the light cone is $|m-n| \simeq 2b|t|$, or equivalently $\rho(m,n) \simeq 2|t|$. The definition~\eqref{eq:rho2} is the same statement written bond by bond for an inhomogeneous chain.

\section{Lieb--Robinson bound in the transport metric} \label{sec:LR}

The purpose of this section is to prove a propagation bound on the Jacobi chain. Specifically, we derive a Lieb--Robinson bound~\cite{LiebRobinson1972,Hastings2006,Nachtergaele2006}, but written in the distance~\eqref{eq:rho2}. We adapt the standard Lieb--Robinson theory, originally developed for locally interacting lattice systems, to the nonuniform nearest-neighbor Jacobi chain, where the natural metric is the transport distance rather than the graph distance. We also explain why the bound implies the velocity $2$.

Let $J_M$ be a finite Jacobi matrix on the sites $\{0,1,\dots,M\}$. For $\lambda>0$ define
\begin{equation}
   \Omega_{M,\lambda} = \max_{0\le r\le M} \left( \mathbf{1}_{r>0}\,b_{r-1}\sinh\left(\frac{\lambda}{b_{r-1}}\right) + \mathbf{1}_{r<M}\,b_r\sinh\left(\frac{\lambda}{b_r}\right) \right). \label{eq:omegadef}
\end{equation}
Here the indicator function, such as $\mathbf{1}_{r>0}$, is $1$ if the specified condition is satisfied and $0$ otherwise.

\begin{theorem}[Finite Krylov Lieb--Robinson bound]\label{thm:LR}
For all $m,n\in\{0,\dots,M\}$, all $t \in \R$, and all $\lambda > 0$,
\begin{equation}
   |\langle m|\e^{-\ii tJ_M}|n\rangle| \le \exp\{-\lambda\rho(m,n)+\Omega_{M,\lambda}|t|\}. \label{eq:LRbound}
\end{equation}
If $0 < \inf_j b_j \le \sup_j b_j < \infty$, then
\begin{equation}
   \lim_{\lambda \to 0} \frac{\Omega_{M,\lambda}}{\lambda} = 2.
\end{equation}
Thus the Lieb--Robinson velocity in the $\rho$-metric is $\vlr = 2$.
\end{theorem}

\noindent
The parameter $\lambda$ should not be confused with a physical velocity. It is an auxiliary parameter controlling the strength of the exponential weight used in the proof. For each fixed $\lambda > 0$, the bound~\eqref{eq:LRbound} can be written as
\begin{equation}
   |\langle m|\e^{-\ii tJ_M}|n\rangle| \le \exp[-\lambda(\rho(m,n)-v_\lambda|t|)],\qquad v_\lambda = \Omega_{M,\lambda}/\lambda. \label{eq:conerewrite}
\end{equation}
Thus $v_\lambda$ is an effective velocity associated with this particular exponential weight. 
The Lieb--Robinson velocity is
the infimum velocity in this family of bounds, which is obtained in
the limit $\lambda\to0^+$. In this limit,
\begin{equation}
   b_j\sinh(\lambda/b_j) = \lambda + O(\lambda^3), \label{eq:sinhexp}
\end{equation}
and the two nearest-neighbor bonds adjacent to a bulk site contribute one $\lambda$ each. Hence $\Omega_{M,\lambda} = 2\lambda + O(\lambda^3)$, and the Krylov-space Lieb--Robinson velocity in the transport metric is
\begin{equation}
   \vlr = \lim_{\lambda \to 0} \frac{\Omega_{M,\lambda}}{\lambda} = 2.
\end{equation}
Finite values of $\lambda$ are useful for bounding the deep exponential tail outside the light cone, but they contain $O(\lambda^2)$ curvature corrections and therefore do not define the asymptotic front velocity.

A proof is given in Appendix~\ref{app:LRproof}; here we indicate the main idea. Fix the source site $n$ and conjugate by the exponential weight
\begin{equation}
   W_\lambda|k\rangle = \e^{\lambda\rho(k,n)}|k\rangle,\qquad K_\lambda = W_\lambda J_M W_\lambda^{-1}.
\end{equation}
Then
\begin{equation}
   \langle m|\e^{-\ii tJ_M}|n\rangle = \e^{-\lambda\rho(m,n)}\langle m|\e^{-\ii tK_\lambda}|n\rangle.
\end{equation}
The non-self-adjoint part of $K_\lambda$ has edge magnitude $b_j\sinh(\lambda/b_j)$. A Schur bound gives $\|\mathrm{Im}\,K_\lambda\| \le \Omega_{M,\lambda}$, and Gr\"onwall's inequality gives $\|\e^{-\ii tK_\lambda}\| \le \e^{\Omega_{M,\lambda}|t|}$. Combining these estimates proves~\eqref{eq:LRbound}.

\section{Round-trip control of the return-amplitude error} \label{sec:roundtrip}

The Lieb--Robinson bound is a one-way bound. The truncation error of the return amplitude is not a one-way observable. It is affected only after information reaches the boundary of the truncated Krylov chain and returns to the initial site. This section turns that idea into an estimate.

Let $J$ be the full Jacobi operator and cut the boundary edge $M\leftrightarrow M+1$:
\begin{equation}
   J = \widehat{J}_M + V_M,\qquad V_M = b_M\left(|M+1\rangle\langle M|+|M\rangle\langle M+1|\right).
\end{equation}
Here $\widehat{J}_M$ decouples the left block $\{0,\dots,M\}$ from the right block. This block has dimension $M+1$, so it realizes the principal truncation $J_{M+1}$ of Section~\ref{sec:jacobi}; in the break-time notation of Definition~\ref{def:breaktime} the cut at edge $M \leftrightarrow M+1$ corresponds to $m = M+1$. The one-step mismatch $\tau_{M+1} - \tau_M = 1/b_M$ between the two labels is of the same order as, and absorbed into, the sublinear correction obtained below, so it does not affect the slope-one statement. Duhamel's formula gives
\begin{equation}
   \e^{-\ii tJ}-\e^{-\ii t\widehat{J}_M} = -\ii\int_0^t\e^{-\ii(t-s)J}V_M\e^{-\ii s\widehat{J}_M}\,\dd s.
\end{equation}
Taking the $(0,0)$ matrix element, the block structure of $\widehat{J}_M$ leaves only the term that crosses the cut once. With
\begin{equation}
   \varDelta A_M(t) \coloneqq \langle 0 | \e^{-\ii tJ} | 0 \rangle - \langle 0 | \e^{-\ii t\widehat{J}_M} | 0 \rangle,
\end{equation}
we obtain
\begin{equation}
   |\varDelta A_M(t)| \le b_M\int_0^t|\langle0|\e^{-\ii(t-s)J}|M+1\rangle|\,|\langle M|\e^{-\ii s\widehat{J}_M}|0\rangle|\,\dd s.
\end{equation}
In this formula each factor is a one-way propagator. The first factor involves the full operator $J$, while Theorem~\ref{thm:LR} is stated for a finite Jacobi matrix. To apply the bound uniformly, set
\begin{equation}
   \Omega_\lambda \coloneqq \sup_M \Omega_{M,\lambda},
\end{equation}
which is finite for bounded Jacobi profiles with $\inf_j b_j > 0$, since $b_j\sinh(\lambda/b_j)$ is then uniformly bounded (see Appendix~\ref{app:LRproof}). With this uniform rate, Theorem~\ref{thm:LR} holds with $\Omega_{M,\lambda}$ replaced by $\Omega_\lambda$ for every finite truncation, and therefore also for the full semi-infinite chain, since the finite-chain bound is uniform in the truncation length and survives as the chain is extended to infinity. Applying it to both one-way factors gives
\begin{equation}
   |\varDelta A_M(t)| \le b_M t\exp\{-\lambda(\rho_M+\rho_{M+1})+\Omega_\lambda t\}, \label{eq:roundtripbound}
\end{equation}
where $\rho_M = \rho(0,M) = \tau_M$.

The exponent contains the round-trip distance
\begin{equation}
   \rho_M + \rho_{M+1} = 2\tau_M + \frac{1}{b_M}.
\end{equation}
In the small-$\lambda$ regime, $\Omega_\lambda/\lambda \to 2$. Thus the distance $2\tau_M$ divided by the velocity $2$ gives the scale $\tau_M$. The break time is therefore $\tau_M$ rather than $2\tau_M$.

\begin{proposition}[Break-time lower bound]\label{prop:lower}
Let $t_\ast(M;\epsilon)$ be defined by~\eqref{eq:breaktime}. For every $\lambda>0$,
\begin{equation}
   \Omega_\lambda t_\ast(M;\epsilon) \ge \lambda(\rho_M+\rho_{M+1}) - \log\left(\frac{b_M t_\ast(M;\epsilon)}{\epsilon}\right). \label{eq:proplower}
\end{equation}
Consequently, for fixed $\epsilon$ and bounded Jacobi profiles with $\inf_j b_j>0$,
\begin{equation}
   \liminf_{M \to \infty} \frac{t_\ast(M;\epsilon)}{\tau_M} \ge 1.
\end{equation}
\end{proposition}

\noindent
The logarithmic term in~\eqref{eq:proplower} is subleading compared with $\tau_M$ when $\tau_M \to \infty$. A more precise optimization gives the Airy-edge correction
\begin{equation}
   t_\ast(M;\epsilon) \ge \tau_M - \left(\frac{9}{32\,b^2}\right)^{1/3}\tau_M^{1/3}\left(\log\left(\frac{b_M\tau_M}{\epsilon}\right)\right)^{2/3},\qquad b=\inf_j b_j > 0. \label{eq:airy}
\end{equation}
The derivation is given in Appendix~\ref{app:refinedlb}. For the main text, the point is that the correction is sublinear in $\tau_M$~\footnote{Sublinear here means it grows only like $\tau_M^{1/3}$, much more slowly than $\tau_M$ itself, so for long chains it is negligible next to the leading term and the reliable window is still set by $\tau_M$.}; the precise constant $(9/32)^{1/3}$ comes from optimizing the exponential-weight parameter and is not essential to the slope-one statement.

It is also useful to distinguish the return-amplitude break time from a one-way arrival time. For a fixed small threshold $\epsilon_{\mathrm{arr}}$, define
\begin{equation}
   t_{\mathrm{arr}}(m;\epsilon_{\mathrm{arr}}) = \inf\{ t \ge 0 \colon |\langle m|\e^{-\ii tJ}|0\rangle|^2 > \epsilon_{\mathrm{arr}} \}.
\end{equation}
Since this is a one-way observable, the same velocity $2$ predicts $t_{\mathrm{arr}} \simeq \tau_m/2$. The return-amplitude error instead gives $t_\ast \simeq \tau_m$.

\section{Saturation and spectral capture} \label{sec:saturation}

The previous section proves that the Krylov approximation cannot fail before the round-trip light cone closes, up to sublinear corrections. This is only a lower bound: it guarantees a reliable window of length $\tau_m$, but says nothing about what happens once the cone has closed. The aim of this section is to describe those later times, and the main message is that two very different things can happen. In the first case there is genuine transport: some amplitude actually travels out to the truncation boundary and back, the lower bound is saturated, and the approximation does start to fail around $t \simeq \tau_m$. In the second case, called \emph{spectral capture}, the finite Krylov block has already resolved all the spectral weight visible from the probe; then essentially no amplitude ever reaches the boundary, the truncation error stays below tolerance at all times, and the break time is effectively infinite. Deciding which case occurs requires spectral information about the Jacobi operator, so we take the two cases in turn.

\subsection{Saturation: the transport-limited regime}

Consider first genuine transport, where the lower bound is saturated. The upper bound (that $\tau_m$ is also an upper bound on $t_\ast$, so the approximation really does fail around $t \simeq \tau_m$) holds in three classes of Jacobi chain, all treated in detail in Appendix~\ref{app:refined} (Proposition~\ref{prop:saturation}): the free (homogeneous) chain, its trace-class perturbations, and broadband absolutely continuous chains. In each, the relevant spectral component actually transports; this is the extra spectral input that the Lieb--Robinson bound alone does not supply. For the free half-line Jacobi chain, with $\alpha_j = \alpha_\infty$ and $b_j = b_\infty$, the propagator is expressed in terms of Bessel functions. The front has the usual Airy scaling at the turning point, and one obtains
\begin{equation}
   t_\ast(M;\epsilon) = \tau_M - \Theta\left(\tau_M^{1/3}(\log(1/\epsilon))^{2/3}\right),\qquad \tau_M = \frac{M}{b_\infty}. \label{eq:airy_scaling}
\end{equation}
The notation $\Theta(\cdot)$ should be read as a two-sided estimate: the correction is of this order both from above and from below, up to constant factors. Equation~\eqref{eq:airy_scaling} therefore gives more than the statement that the correction is small compared with $\tau_M$; it identifies the leading Airy-front scale $\tau_M^{1/3}(\log(1/\epsilon))^{2/3}$. This scale is sublinear, so $t_\ast(M;\epsilon)/\tau_M \to 1$, but it is the dominant finite-$M$ shift of the break time. The asymptotic slope is one.

Trace-class perturbations of the free chain also preserve the ballistic front up to an $O(1)$ scattering delay, and broadband absolutely continuous Jacobi chains give the same slope-one behavior under standard assumptions; see Appendix~\ref{app:refined}. 
The remaining possibility is the one in which this upper bound \emph{fails}: when the relevant component does not transport but is instead localized or already resolved, the break time is infinite rather than $\simeq\tau_m$. We turn to that case, spectral capture, now.

\subsection{Spectral capture}

Spectral capture is the opposite situation. Here the right question is not how fast information reaches the boundary, but whether any information needs to reach the boundary at all. The probe overlaps with the eigenstates $|E_\ell\rangle$ of $H$ with weights $w_\ell=|\langle E_\ell|v_0\rangle|^2$. Suppose the $m$-dimensional truncation has resolved these weights except for a residual
\begin{equation}
   \eta_m = \sum_{\ell \colon \text{unresolved}} w_\ell,
\end{equation}
the total weight of the spectral lines not yet captured by the Krylov chain. The $m$-dimensional chain then already reproduces the exact amplitude except for these still-unresolved lines; if their total weight is already below the tolerance, the approximation is accurate at \emph{all} times, and nothing is left for the light cone to carry out to the boundary. Hence~\footnote{In measure form, $A(t) = \int\e^{-\ii Et}\dd\mu$ with the probe's local spectral measure $\dd\mu(E) = \sum_\ell w_\ell\,\delta(E-E_\ell)\,\dd E$, and $A_m(t) = \int\e^{-\ii Et}\dd\mu_m$ with $\mu_m$ the spectral measure of the truncated chain $J_m$ (the weights $|\langle 0|E^{(m)}_\ell\rangle|^2$ of the eigenvalues $E^{(m)}_\ell$ of $J_m$). These two probability measures share a common resolved part $c$ and differ only by residuals $\nu$, $\nu_m$ of mass $\nu(\R) = \nu_m(\R) = \eta_m$: writing $\mu = c + \nu$ and $\mu_m = c + \nu_m$, the common part cancels and $|A(t)-A_m(t)| = \big|\int\e^{-\ii Et}(\dd \nu-\dd \nu_m)\big|\le \nu(\R)+\nu_m(\R) = 2\eta_m$. The factor of two counts the two residuals.}
\begin{equation}
   |A(t)-A_m(t)| \le 2 \eta_m \qquad \text{for all }t. \label{eq:capture}
\end{equation}

Therefore, once $2\eta_m < \epsilon$, the break time at tolerance $\epsilon$ is infinite in the ideal noiseless model:
\begin{equation}
   t_\ast(m;\epsilon) = \infty.
\end{equation}
This gives two regimes,
\begin{align}
   &\text{transport-limited:} && m  <  m_\ast(\epsilon),\quad t_\ast(m) \simeq \tau_m, \\
   &\text{spectral capture:}  && m \ge m_\ast(\epsilon),\quad t_\ast(m) = \infty.
\end{align}
Here $m_\ast(\epsilon)$ is fixed by the residual weight alone,
\begin{equation}
   m_\ast(\epsilon) = \min\{ m \colon 2\eta_m \le \epsilon \}, \label{eq:mstar}
\end{equation}
so $m_\ast(\epsilon)$ is the number of Lanczos steps needed to resolve the spectral measure to within the tolerance: below it the break time grows with transport, $t_\ast \simeq \tau_m$, while above it the truncation is already accurate and does not break. Because $m_\ast$ is set by how the initial state spreads its weight over the spectrum, it depends strongly on the probe: a state with narrow spectral support is captured after only a few Lanczos steps, whereas a broad-spectrum state stays transport-limited over a wide range of $m$. This state dependence is shown in Fig.~\ref{fig:capture} of Section~\ref{sec:numerics}. It is explicit in the two extreme cases. If the cyclic subspace $\KK(H,v_0)$ is finite, of dimension $d_B$, the Jacobi chain terminates ($b_{d_B} = 0$) and the truncation becomes exact at $m = d_B$; then $m_\ast = d_B$ and the capture is sharp and $\epsilon$-independent. 
For example, in the $N$-site uniform ring studied numerically in Section~\ref{sec:numerics}, reflection symmetry protects the odd-parity modes as a dark sector (Appendix~\ref{app:spectral}), yielding $d_B = \lfloor N/2\rfloor+1$.
If instead the important part of the state is localized over a length $\xi$, then the amplitude that leaks out to the truncation boundary falls off exponentially with distance, $O(\e^{-m/\xi})$, so the unresolved weight shrinks as $\eta_m \sim \e^{-m/\xi}$. Only about $\xi\log(1/\epsilon)$ Lanczos steps are then needed to push it below the tolerance, essentially independent of how long the chain is, and the state is captured early (see Appendix~\ref{app:refined}). At the opposite extreme, a state spread over a broad continuous spectrum never terminates the chain and its unresolved weight shrinks only slowly; then $m_\ast$ grows as the tolerance is tightened, and in practice one always stays in the transport-limited regime, where $t_\ast \simeq \tau_m$ governs the whole accessible window.

This second regime is not a violation of the Lieb--Robinson picture; it is a limiting case of it. The Lieb--Robinson bound controls transport of boundary errors, but when the boundary contribution is already below the tolerance there is simply no error left to transport. If the relevant spectral measure has already been resolved, the boundary contribution is below the tolerance for all times. In localized pure-point components this can happen because boundary amplitudes are exponentially small in the distance. In finite-dimensional or nearly finite-dimensional bright sectors (the subspace of eigenstates visible to the probe, see Appendix~\ref{app:spectral}) it happens because the Krylov subspace has already captured all observable spectral lines.

\section{Numerical evidence} \label{sec:numerics}

This section summarizes the numerical checks. The aim is not to prove universality from numerics, but to verify the two specific predictions above: the transport-limited slope $t_\ast \simeq \tau_m$, and the distinction between one-way arrival and round-trip return.

\begin{figure}[p]
\centering
\includegraphics[width=\linewidth]{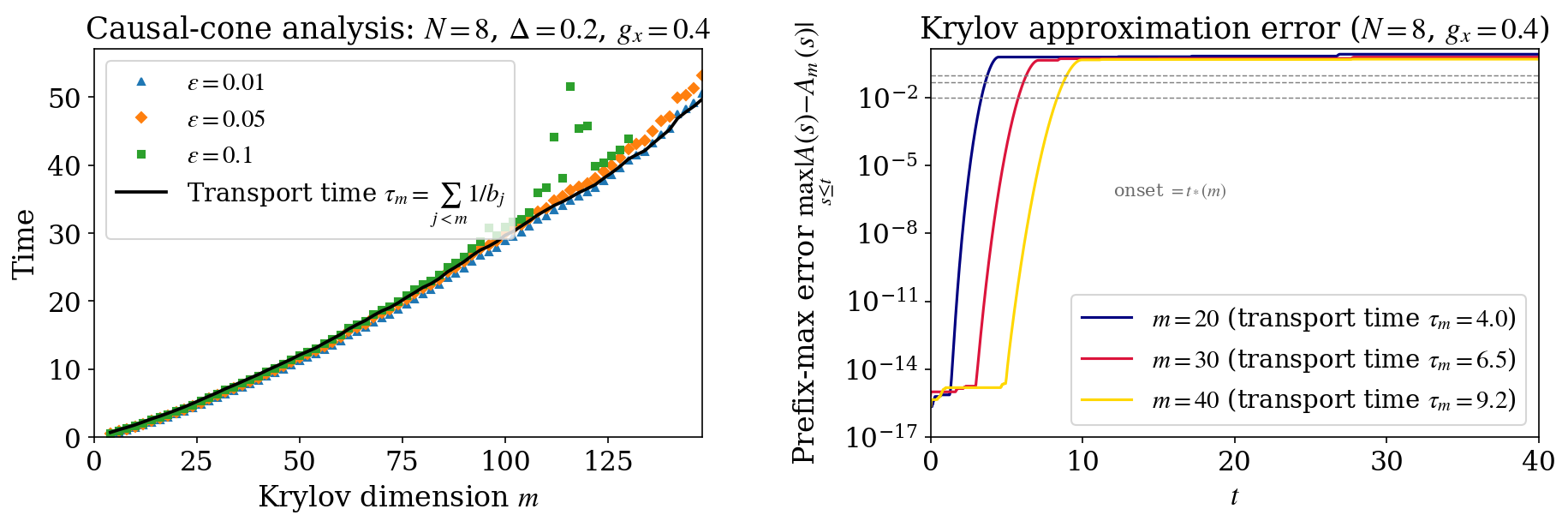}
\caption{Krylov causal cone and the definition of the break time. The break time $t_\ast(m;\epsilon)$ is the first time at which the prefix-maximum error $\max_{s \le t}|A(s)-A_m(s)|$ exceeds the tolerance $\epsilon$. Left: for the $N = 8$ XXZ ring with $\Delta = 0.2$ and $g_x = 0.4$ and a broad-spectrum probe, $t_\ast(m;\epsilon)$ tracks the transport time $\tau_m = \sum_{j<m} 1/b_j$ with a slope close to one for all three tolerances. A broad probe excites essentially the whole spectrum, so it stays transport-limited over the entire accessible range and does not undergo spectral capture until the chain itself terminates; the state-dependent capture dimension $m_\ast$ and its dependence on the initial state are shown separately in Fig.~\ref{fig:capture}. Right: for fixed $m$, the error stays small until a sharp onset near $t \simeq \tau_m$; this onset is the break time $t_\ast(m)$ itself (a time, not a Krylov dimension). This is the numerical signature of the Lieb--Robinson light cone in the inhomogeneous Krylov metric.}
\label{fig:causalcone}
\end{figure}

The main spin-chain example is the XXZ ring with transverse field,
\begin{equation}
   H = \sum_{j=1}^N h_j\,( X_j X_{j+1} + Y_j Y_{j+1} + \Delta Z_j Z_{j+1} ) + g_x \sum_{j=1}^N X_j,\qquad j+N \equiv j,
\end{equation}
where $h_j$ is the exchange coupling, $\Delta$ the anisotropy, $g_x$ the transverse field, and $X_j$, $Y_j$, $Z_j$ are the Pauli matrices at site $j$. Unless otherwise stated, $h_j = 1$ and $\Delta = 0.2$. For $N = 8$ and $g_x = 0.4$, Fig.~\ref{fig:causalcone} shows that the break time defined by~\eqref{eq:breaktime} follows $\tau_m$ with a slope close to one in the transport-limited region. The error curves show a sharp onset near $t \simeq \tau_m$, which is the expected light-cone behavior.

\begin{figure}[p]
\centering
\includegraphics[width=\linewidth]{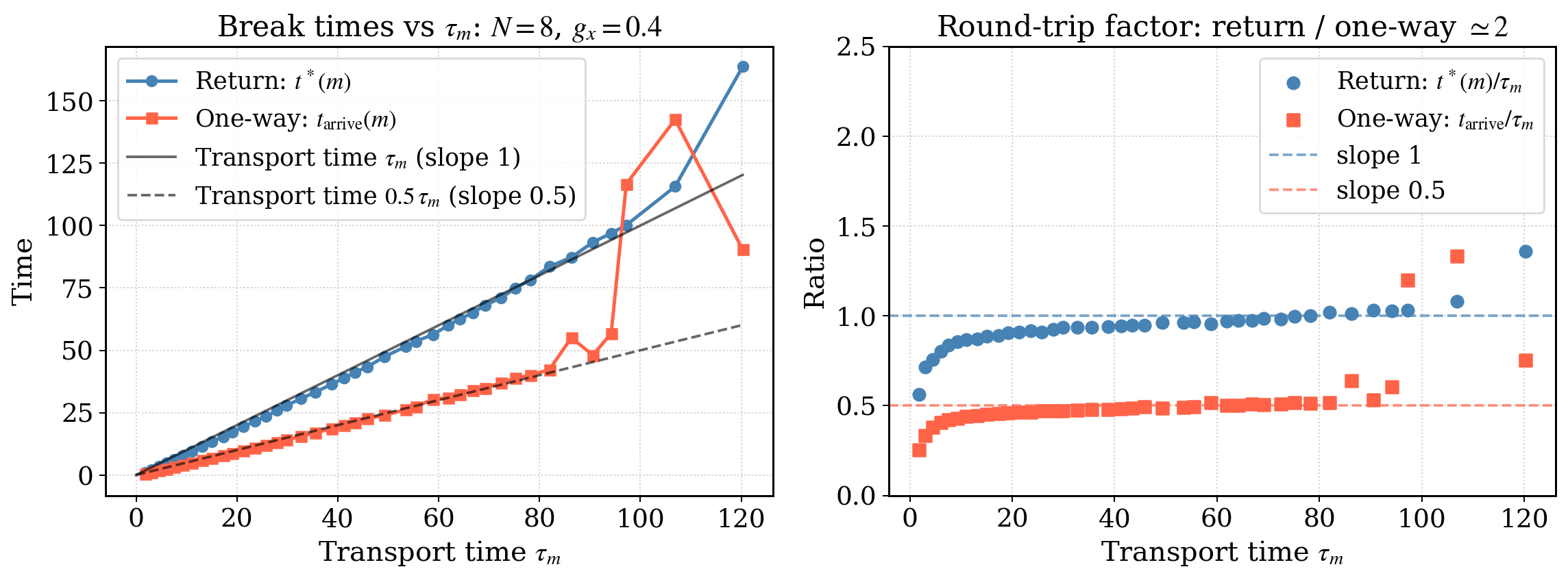}
\caption{One-way arrival and round-trip break time measure the same light cone. The red data show the endpoint arrival time $t_{\mathrm{arr}}$, defined by the first appearance of probability at the Krylov boundary. It scales as $t_{\mathrm{arr}}/\tau_m \simeq 1/2$, as expected for one-way propagation with velocity $2$. The blue data show the return-amplitude break time $t_\ast$, which scales as $t_\ast/\tau_m \simeq 1$ because the boundary signal must travel out and back. The two slopes therefore support the same Lieb--Robinson velocity, not two different velocities.}
\label{fig:roundtrip}
\end{figure}

The factor-of-two interpretation is checked in Fig.~\ref{fig:roundtrip}. The endpoint arrival time scales as $\tau_m/2$, while the return-amplitude break time scales as $\tau_m$. This is a useful test because both quantities are computed from the same Jacobi dynamics but probe different paths. The result supports a single velocity $\vlr = 2$.

\begin{figure}[p]
\centering
\includegraphics[width=\linewidth]{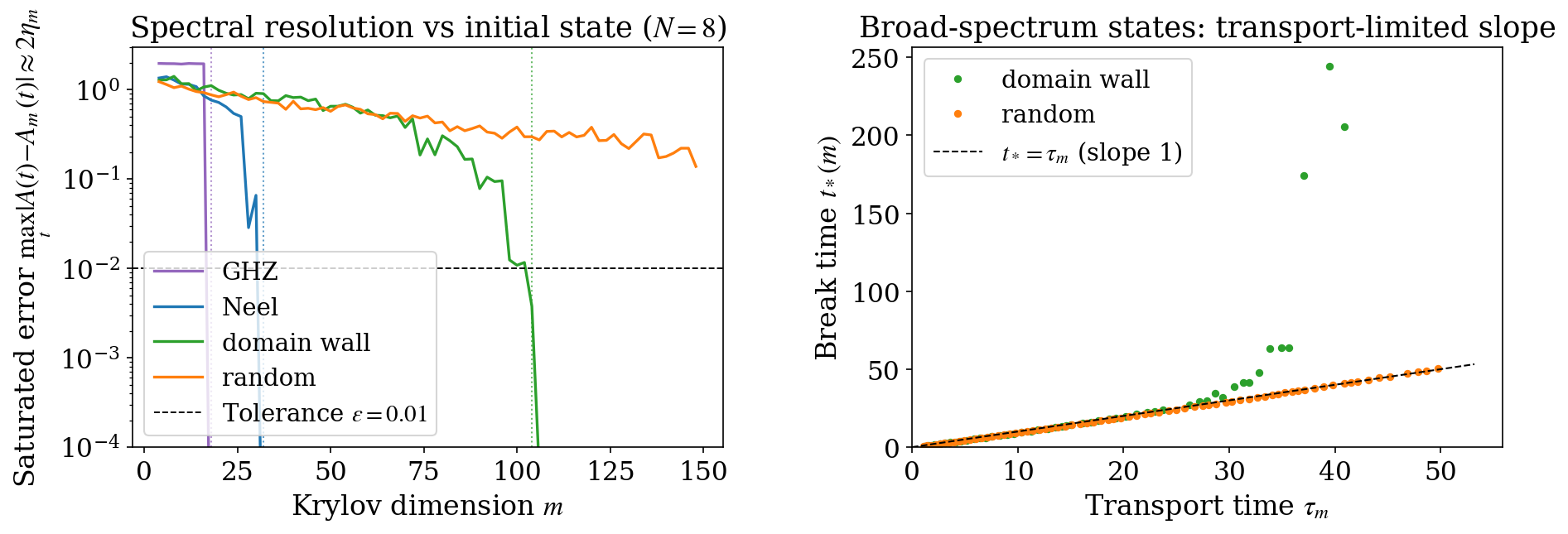}
\caption{Initial-state dependence of spectral capture. Left: the saturated amplitude error $\max_t|A(t)-A_m(t)| \approx 2\eta_m$ as a function of Krylov dimension $m$, for four initial states of the $N = 8$ XXZ ring (Greenberger--Horne--Zeilinger (GHZ), N\'eel, domain wall, and a random broad-spectrum state). Each state resolves its spectral measure at a different rate, so the error drops below the tolerance $\epsilon$ (dashed line) at a different, state-dependent dimension $m_\ast$: a narrow-spectrum state such as GHZ is captured after only a few steps ($m_\ast \approx 18$), whereas a broad state stays uncaptured over the whole accessible range. Right: for the broad-spectrum states the break time $t_\ast(m)$ still tracks $\tau_m$ with slope one until capture, confirming that $m_\ast$ shifts with the initial state while the transport law $t_\ast \simeq \tau_m$ is universal. The capture jump is caused by spectral resolution, not by any change of the Lieb--Robinson velocity.}
\label{fig:capture}
\end{figure}

\begin{figure}[p]
\centering
\includegraphics[width=\linewidth]{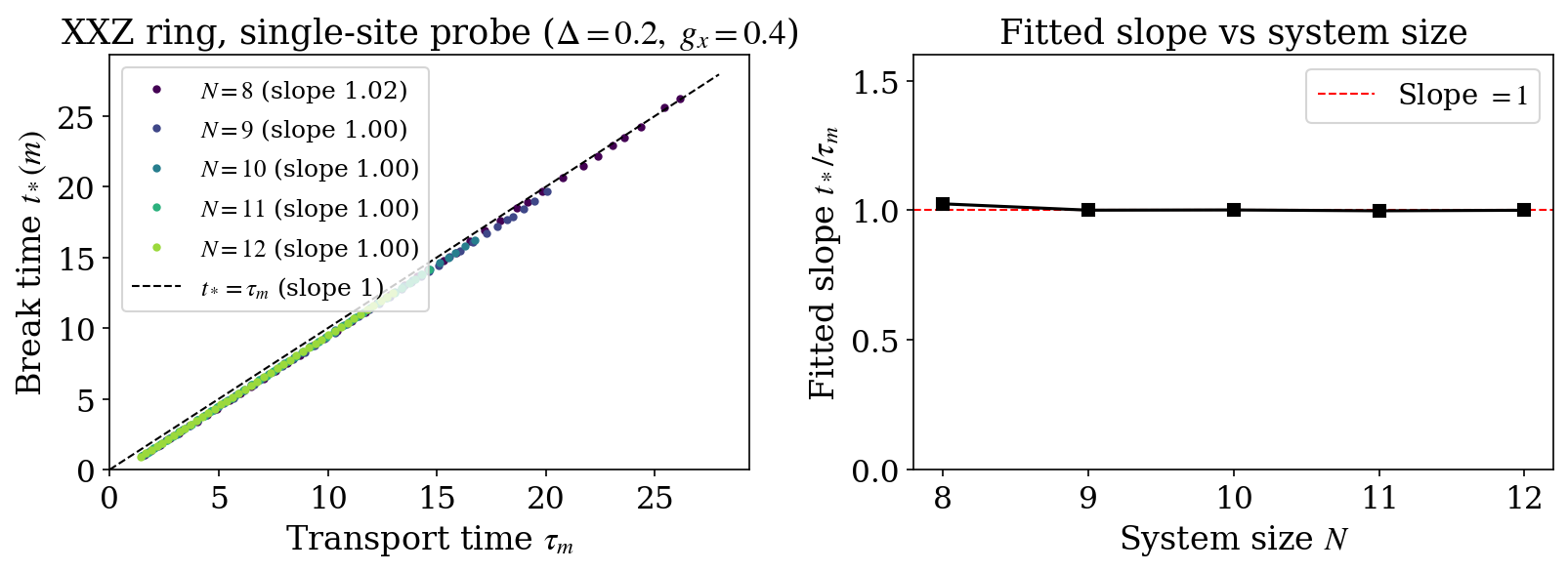}
\caption{Transport-limited slope for the XXZ ring at several system sizes (single-site probe $|v_0\rangle = X_{N/2}|0\cdots0\rangle$, $\Delta = 0.2$, $g_x = 0.4$, tolerance $\epsilon = 10^{-2}$). Left: the break time $t_\ast(m)$ against the transport time $\tau_m$ for $N = 8$--$12$; all sizes track $t_\ast \simeq \tau_m$. Right: the fitted slope stays close to one for every $N$ (even sizes and $N \ge 9$), so the transport law is not a small-size artifact.}
\label{fig:sizes}
\end{figure}

\begin{figure}[p]
\centering
\includegraphics[width=0.5\linewidth]{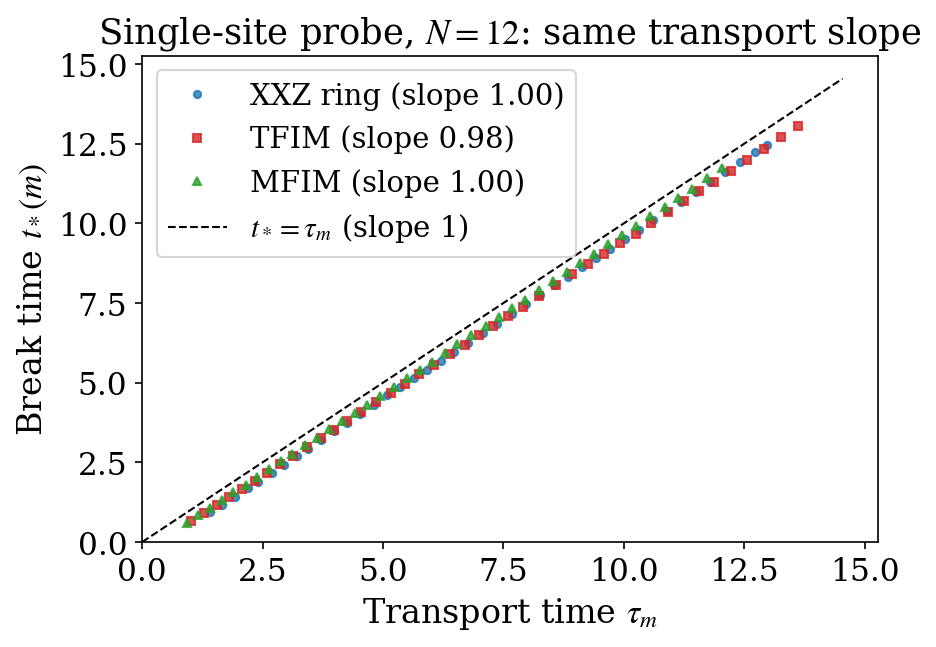}
\caption{Same transport slope for three models (single-site probe, $N = 12$, $\epsilon = 10^{-2}$): the XXZ ring, the transverse-field Ising model (TFIM, $\sum_j Z_j Z_{j+1} + \sum_j X_j$), and the mixed-field Ising model (MFIM, $\sum_j Z_j Z_{j+1} + 1.05 \sum_j X_j + 0.5 \sum_j Z_j$). After passing to the transport coordinate $\tau_m$, all three collapse onto $t_\ast \simeq \tau_m$; the model-dependent Lanczos plateaus are absorbed into the metric.}
\label{fig:models}
\end{figure}

The slope-one behavior was also tested for different initial states (see Fig.~\ref{fig:capture}). For the XXZ ring, even-numbered system sizes and $N \ge 9$ give slopes close to one for single-site probes (Fig.~\ref{fig:sizes}). Very small systems can give inflated slopes because the transport-limited fitting window is short. For several initial states at $N = 8$, broad-spectrum states collapse well onto $t_\ast \simeq \tau_m$, while narrow-spectrum states enter spectral capture earlier. Comparisons with transverse-field Ising and mixed-field Ising models show the same trend (Fig.~\ref{fig:models}): after using the transport coordinate $\tau_m$, model-dependent plateau values of the Lanczos coefficients are largely absorbed into the metric.

\begin{figure}[p]
\centering
\includegraphics[width=\linewidth]{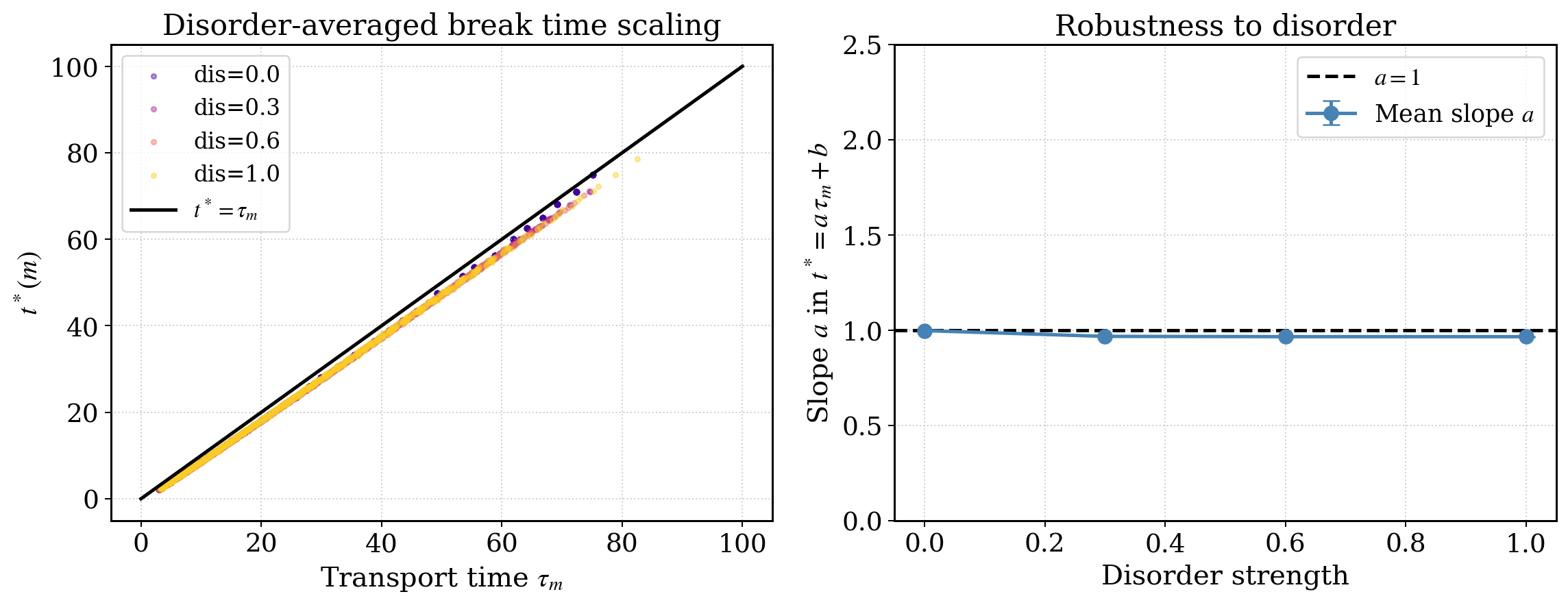}
\caption{Robustness of the break time scaling against spatial disorder. The break time scaling is tested on the XXZ ring ($N = 8$, $g_x = 0.4$) with disordered exchange couplings $h_j = 1 + \sigma \zeta_j$, where $\zeta_j \in [-1,1]$ and the disorder strength is $\sigma \in \{0.0, 0.3, 0.6, 1.0\}$. The data at each $\sigma$ is averaged over $20$ disorder realizations. Left: the disorder-averaged break time $t_\ast(m)$ versus the transport time $\tau_m$. The linear relationship remains stable. Right: the fitted slope $a$ as a function of the disorder strength $\sigma$. The slope stays close to $1$, i.e., $a = 0.997, 0.967, 0.965, 0.965$ for $\sigma = 0.0, 0.3, 0.6, 1.0$, respectively. This confirms that the transport-limited scaling $t_\ast \simeq \tau_m$ is robust against disorder and is not an artifact of translation-invariant couplings.}
\label{fig:disorder}
\end{figure}

Figure~\ref{fig:disorder} shows the disorder tests ($h_j = 1 + \sigma \zeta_j$, $\zeta_j \in[-1,1]$, $N = 8$, $g_x = 0.4$) for the disorder strength $\sigma \in \{0.0, 0.3, 0.6, 1.0\}$, averaging the fitted slope over $20$ disorder realizations at each $\sigma$; the fitted slope stays close to one throughout, $a = 0.997, 0.967, 0.965, 0.965$ for $\sigma = 0.0, 0.3, 0.6, 1.0$, respectively, confirming that the transport-limited slope is not an artifact of the translation-invariant couplings.

\begin{figure}[p]
\centering
\includegraphics[width=\linewidth]{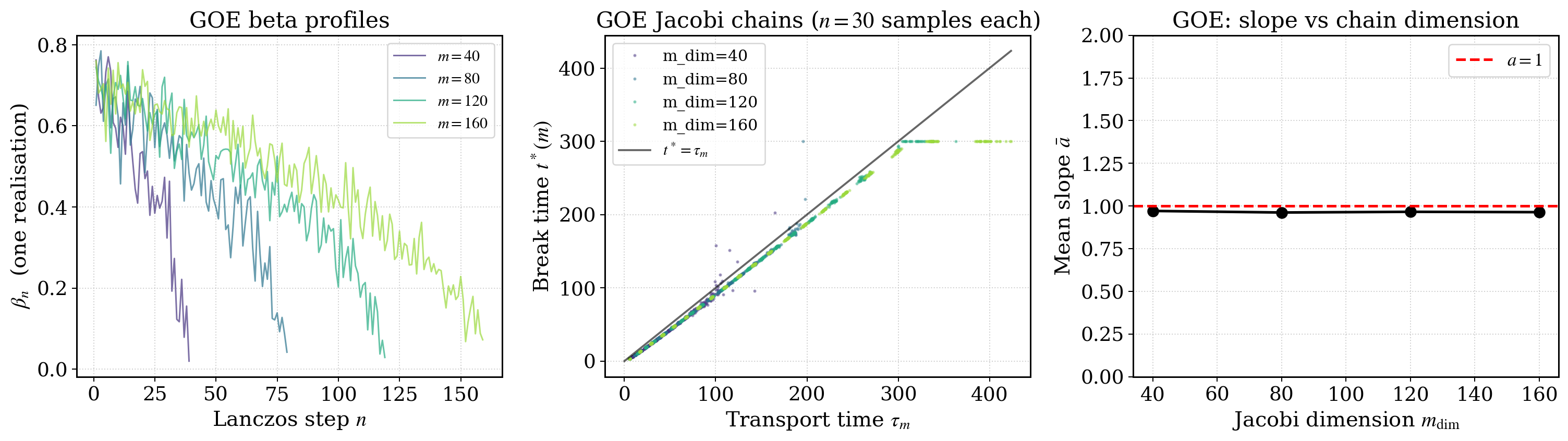}
\caption{Random Jacobi-chain test without a spin-chain background. The Dumitriu--Edelman tridiagonal GOE ensemble gives Jacobi matrices directly. The center panel shows that the measured break times collapse onto the line $t_\ast = \tau_m$ for many realizations and truncation depths. The right panel shows fitted slopes close to one as the matrix dimension is varied. This supports the interpretation that the transport formula is a property of the nearest-neighbor Jacobi chain in the metric $\rho$, not of a particular microscopic model.}
\label{fig:goe}
\end{figure}

A further test removes the physical spin lattice entirely (Fig.~\ref{fig:goe}). We sample tridiagonal Gaussian Orthogonal Ensemble (GOE) Jacobi matrices in the Dumitriu--Edelman form~\cite{DumitriuEdelman2002}. The initial vector is $|0\rangle$, and break times are measured for principal truncations. For matrix dimensions 40--160, fitted slopes lie near one and show no systematic drift away from it. This indicates that the mechanism is Jacobi-chain transport, rather than a peculiarity of the XXZ Hamiltonian.

A few conventions are used throughout. The break time is always the prefix-maximum quantity of Eq.~\eqref{eq:breaktime}, so a time $t$ counts as reliable only if the error stays below threshold on all of $[0,t]$; the linear fits of $t_\ast$ against $\tau_m$ use only points before spectral capture, and points whose error never crosses the threshold within the observation window are not counted as finite break-time data. For the XXZ ring the parameters are $h_j = 1$, $\Delta = 0.2$, $g_x = 0.4$, with disorder tests $h_j = 1 + \sigma \zeta_j$ ($\sigma$ is the disorder strength and $\zeta_j \in [-1,1]$); the model comparisons use the transverse-field and mixed-field Ising chains, and the random-Jacobi test uses the Dumitriu--Edelman tridiagonal GOE ensemble (Gaussian diagonal, off-diagonal entries scaled to the Wigner semicircle). Throughout, $b_j$ denotes the hopping on the edge $j \leftrightarrow j+1$, which avoids the usual shift ambiguity in the Lanczos $\beta_n$.

\clearpage
\section{Implications for Krylov simulation and Hamiltonian learning} \label{sec:discussion}

\subsection{How large a Krylov dimension is needed}

This section addresses, in practical terms, how many Lanczos steps are enough. To reproduce the survival amplitude reliably up to a target time $T$, one should run the recursion until the transport time reaches $T$; the required Krylov dimension is
\begin{equation}
   m_{\mathrm{req}}(T) = \min\left\{ m \colon \tau_m = \sum_{j<m} \frac{1}{b_j} \ge T \right\}. \label{eq:mreq}
\end{equation}
Because $\tau_m$ is assembled only from the hoppings $b_j$, which the Lanczos recursion produces one at a time, $m_{\mathrm{req}}$ is available on the fly at no extra cost: the horizon can be read off from the coefficients themselves, without a separate error estimator. Each additional Krylov vector extends the reliable window by about
\begin{equation}
   \varDelta t_\ast(m) \coloneqq t_\ast(m+1) - t_\ast(m) \simeq \frac{1}{b_m},
\end{equation}
so a small hopping at the current frontier means one more step buys a long time, while a large hopping means diminishing returns. One may stop before $m_{\mathrm{req}}$ if spectral capture has already set in ($2\eta_m \le \epsilon$), which is equally visible from the computed spectrum. In the ballistic (transport-limited) case the estimate $m \simeq m_{\mathrm{req}}$ is asymptotically tight, and the sublinear correction of Eq.~\eqref{eq:airy} sharpens it when a certified tolerance is required.

The result should not be read as saying that the $m$-dimensional Krylov approximation always fails exactly at $t = \tau_m$, for every Krylov dimension $m$ and every initial state $|v_0\rangle$: the bound is one-sided, and saturation requires genuine transport in the relevant spectral component. When the observable spectral measure has already been resolved, or the remaining component is localized, the error stays below tolerance for all times. This is the spectral-capture regime of Section~\ref{sec:saturation}.

The assumptions should also be kept visible. The lower bound uses bounded Jacobi profiles with a positive lower bound on the hoppings. The half-line version uses the Carleman condition so that infinity is at infinite $\rho$-distance. The saturation statements are proved for particular ballistic classes. Operator Krylov problems with rapidly growing Lanczos coefficients may belong to a different regime and should not be treated by simply importing the present conclusion (for a detailed analysis of the operator Krylov case, see~\cite{matsuura2026opkrylov}).

\subsection{Restricted Hamiltonian learning}

A second application, and the one that motivates this work, is Hamiltonian learning from a single probe: inferring the dynamics of an interacting many-body system from the time series of one controllable spin, as in quantum sensing and boundary spectroscopy on nitrogen-vacancy (NV) center, donor-spin, and quantum-dot spin platforms. The survival amplitude $A(t)$ determines the probe's local spectral measure and, through the orthogonal-polynomial (Favard) correspondence, the entire Jacobi chain $\{\alpha_n,b_n\}$ of the bright cyclic subspace (Appendix~\ref{app:spectral}). A single probe therefore learns exactly this Jacobi model, and no more: couplings in a dark sector, or two microscopic Hamiltonians sharing the same local spectral measure, cannot be distinguished. This is the setting of restricted-access learning~\cite{Burgarth2009}. The break-time result adds a temporal statement: a learned $m$-coefficient chain reproduces the dynamics reliably up to $\tau_m$. The learned model then has a predictive horizon rather than being only a static fit, and this horizon is used in coherent-state import~\cite{matsuura2026hp}.

This state-Krylov picture has a direct operator-Krylov counterpart, developed in a companion paper~\cite{matsuura2026opkrylov}, which removes the main practical obstructions of survival-amplitude learning. Survival-amplitude protocols need a conserved excitation number or an internal reference eigenstate, and so are confined to special symmetries; replacing the survival amplitude by the infinite-temperature autocorrelation $C_\alpha(t) = \langle\mathcal{O}_\alpha(t)\,\mathcal{O}_\alpha(0)\rangle$ of a single probe observable $\mathcal{O}_\alpha$ lifts this restriction. Running Lanczos on the operator (rather than state) Krylov space yields an operator-Krylov Jacobi matrix that can be reconstructed nonparametrically and oracle-free. This reconstruction uses only free evolution under the unknown $H$ together with single-probe preparation and measurement, with no symmetry assumption, no parametric model, and no controlled two-qubit gates, which are conditions natural to the infinite-temperature baths of NV-center and donor-spin devices. The operator-Krylov chain obeys the same Lieb--Robinson bound and transport formula proved here, by the same argument (conjugation by a weighted shift, a finite Schur test, and Gr\"onwall's lemma), with the same velocity $\vlr = 2$ and the same transport-limited-to-spectral-capture dichotomy. The break-time analysis of the present paper therefore carries over and supplies the criterion for how long the autocorrelation must be observed to resolve $m$ chain coefficients, which underlies a sample-complexity guarantee, polynomial in $1/\epsilon$, for learning the spectrum to precision $\epsilon$~\cite{matsuura2026opkrylov}. In this sense the break-time bound established here is the dynamical basis of the restricted-access learning program that motivates this work.

\section{Conclusion} \label{sec:conclusion}

We have described finite-Krylov reliability as a causal propagation problem on the Jacobi chain. The distance which controls the propagation is the inhomogeneous transport metric
\begin{equation}
   \rho(m,n) = \sum_{j=\min(m,n)}^{\max(m,n)-1} \frac{1}{b_j}.
\end{equation}
In this metric, Jacobi dynamics satisfies a Lieb--Robinson bound with velocity $2$. Applying Duhamel's formula to the difference between the full and truncated evolutions shows that the return-amplitude error is controlled by a round trip to the truncation boundary. This gives
\begin{equation}
   t_\ast(m;\epsilon) \ge \tau_m[1-o(1)],\qquad \tau_m = \sum_{j<m} \frac{1}{b_j}.
\end{equation}
The factor of two in the round trip is canceled by the velocity $2$.

This gives a useful interpretation of Lanczos break times: the reliable time window grows as $\tau_m$ until the residual observable spectral weight drops below the tolerance, after which the break time becomes effectively infinite (see Section~\ref{sec:saturation}). The numerical data on spin chains and random Jacobi matrices support this picture.

The practical consequence, given in Section~\ref{sec:discussion}, is to run the Lanczos recursion until $\tau_m$ reaches the target time (stopping earlier under spectral capture); this stopping rule needs only the computed hopping coefficients. The same viewpoint separates what is dynamically controlled by Krylov transport from what is limited by the observable bright sector, which is the basis for the Hamiltonian-learning application.

\appendix

\section{Proof of the Lieb--Robinson bound and the refined lower bound}

This appendix gives the details of Theorem~\ref{thm:LR} in Section~\ref{sec:LR} and of the refined lower bound quoted in Section~\ref{sec:roundtrip}. The proofs are short, but several estimates are easy to misread if written too compactly, so we spell them out.

\subsection{The Lieb--Robinson bound} \label{app:LRproof}

Fix a source site $n$ and define
\begin{equation}
   W_\lambda|k\rangle = \e^{\lambda\rho(k,n)}|k\rangle,\qquad K_\lambda = W_\lambda J_M W_\lambda^{-1}.
\end{equation}
Since the chain is finite, $W_\lambda$ is bounded and invertible. Hence
\begin{equation}
   W_\lambda\e^{-\ii tJ_M} W_\lambda^{-1} = \e^{-\ii tK_\lambda}
\end{equation}
and, because $W_\lambda|n\rangle = |n\rangle$,
\begin{equation}
   \langle m|\e^{-\ii tJ_M}|n\rangle = \e^{-\lambda\rho(m,n)}\langle m|\e^{-\ii tK_\lambda}|n\rangle. \label{eq:conjapp}
\end{equation}
On an edge $j \leftrightarrow j+1$, the conjugation changes the hopping by the ratio of the weights on the two adjacent sites. Along the path away from $n$, this ratio is $\exp(\pm\lambda/b_j)$. Thus the two directed hopping terms on that edge become
\begin{equation}
   b_j\e^{\lambda/b_j}|j+1\rangle\langle j|,\qquad b_j\e^{-\lambda/b_j}|j\rangle\langle j+1|,
\end{equation}
up to orientation. The self-adjoint part contains the $\cosh$ contribution, while the non-self-adjoint part has magnitude
\begin{equation}
   b_j\sinh\left(\frac{\lambda}{b_j}\right)
\end{equation}
on that edge.

We next bound the operator norm (induced $2$-norm) of $\mathrm{Im}\,K_\lambda = (K_\lambda-K_\lambda^\dagger)/(2\ii)$, where the superscript $\dagger$ denotes the conjugate transpose for matrices. We use the elementary Schur estimate: for a general matrix $G = (G_{ij})$, if
\begin{equation}
   R = \sup_i \sum_j |G_{ij}|,\qquad S = \sup_j \sum_i |G_{ij}|,
\end{equation}
then
\begin{equation}
   \|G\| \le \sqrt{RS}.
\end{equation}
In the present tridiagonal case, the row and column sums of $\mathrm{Im}\,K_\lambda$ are bounded by the same number, namely $\Omega_{M,\lambda}$ in~\eqref{eq:omegadef}. Hence
\begin{equation}
   \|\mathrm{Im}\,K_\lambda\| \le \Omega_{M,\lambda}.
\end{equation}
Finally, for an arbitrary state $\psi_0$, let $\psi(t)=\e^{-\ii tK_\lambda}\psi_0$. Then
\begin{align}
   \frac{\dd}{\dd t} \|\psi(t)\|^2
   &= \langle-\ii K_\lambda\psi,\psi\rangle + \langle\psi,-\ii K_\lambda\psi\rangle \nonumber \\
   &= \langle\psi,\ii K_\lambda^\dagger\psi\rangle + \langle\psi,-\ii K_\lambda\psi\rangle \nonumber \\
   &= 2\langle\psi,(\mathrm{Im}\,K_\lambda)\psi\rangle \nonumber \\
   &\le 2\|\mathrm{Im}\,K_\lambda\|\,\|\psi(t)\|^2.
\end{align}
Here we use Gr\"onwall's inequality in the elementary form that if a nonnegative differentiable function $u(t)$ satisfies $\tfrac{\dd}{\dd t} u(t) \le C u(t)$ for some constant $C$, then $u(t) \le u(0) \e^{Ct}$. Applying this to $u(t) = \|\psi(t)\|^2$ with $C = 2\|\mathrm{Im}\,K_\lambda\|$ gives $\| \psi(t) \| \le \e^{ \| \mathrm{Im}\, K_\lambda \| |t|} \| \psi_0 \|$. Then
\begin{equation}
   \|\e^{-\ii tK_\lambda}\| \le \exp(\|\mathrm{Im}\,K_\lambda\|\,|t|) \le \e^{\Omega_{M,\lambda}|t|}. \label{eq:gronwall}
\end{equation}
Combining~\eqref{eq:conjapp} and~\eqref{eq:gronwall} proves~\eqref{eq:LRbound}.

The velocity statement follows from the expansion~\eqref{eq:sinhexp}. Since each row has at most two neighboring bonds,
\begin{equation}
   \Omega_{M,\lambda} = 2\lambda + O(\lambda^3)
\end{equation}
when the hoppings are bounded away from zero. Thus $\Omega_{M,\lambda}/\lambda \to 2$.

There is also a useful Agmon-type version~\cite{Agmon1982,Tran2021}, in which the exponential weight is allowed to vary from edge to edge rather than using a constant slope. Instead of a constant slope $\lambda$ in the distance $\rho$, for an arbitrary parameter $q>0$, choose edge increments
\begin{equation}
   F(j+1) - F(j) = \arsinh\left(\frac{q}{b_j}\right)
\end{equation}
on the path from $n$ to $m$. Then $b_j\sinh(F(j+1)-F(j)) = q$ on every active edge, so the Schur row sum is bounded by $2q$. This gives
\begin{equation}
   |\langle m|\e^{-\ii tJ_M}|n\rangle| \le \exp[-d_q(m,n)+2q|t|],\qquad d_q(m,n) = \sum_j \arsinh\left(\frac{q}{b_j}\right).
\end{equation}
For a homogeneous chain, optimizing over $q$ reproduces the Bessel light-cone threshold $|m-n|=2b|t|$.

For a half-infinite chain, assume the Carleman condition
\begin{equation}
   \sum_{j=0}^\infty \frac{1}{b_j} = \infty.
\end{equation}
This says that infinity is at infinite $\rho$-distance. Under the usual essential self-adjointness assumptions~\cite{ReedSimon2}, finite truncations converge to the half-line operator in the strong resolvent sense. If $\Omega_\lambda = \sup_M\Omega_{M,\lambda} < \infty$, the finite-chain bound passes to the limit by the functional calculus~\cite{ReedSimon2}.

\subsection{Refined lower bound} \label{app:refinedlb}

Here we record the more precise version of the lower bound, the Airy-edge correction quoted in Section~\ref{sec:roundtrip}; it holds for any bounded Jacobi profile with $\inf_j b_j>0$.

Let
\begin{equation}
   b = \inf_j b_j > 0,\qquad B = \sup_j b_j < \infty,\qquad \tau_M = \sum_{j<M} \frac{1}{b_j}.
\end{equation}
Using $\sinh(x) \le x\exp(x^2/6)$,~\footnote{From the Weierstrass product $\sinh(x)/x = \prod_{k \ge 1}(1+x^2/(k^2\pi^2))$ and $1+u \le \e^u$ one gets $\sinh(x)/x \le \exp(x^2\sum_{k \ge 1}1/(k^2\pi^2)) = \e^{x^2/6}$, since $\sum_{k \ge 1}k^{-2} = \pi^2/6$. The bound holds for all real $x$.} we obtain
\begin{equation}
   \Omega_\lambda \le 2\lambda\exp\left(\frac{\lambda^2}{6b^2}\right). \label{eq:curvature}
\end{equation}
This bound keeps the first curvature correction to the linear light cone.

The strategy is to show that the truncation is still reliable up to a time just short of $\tau_M$. Fix a small $\delta > 0$ and set $T_0 = (1-\delta)\tau_M$; we ask how small $\delta$ can be while the round-trip error stays below $\epsilon$ on all of $[0,T_0]$. Take $T_0 = (1-\delta)\tau_M$. If the right-hand side of~\eqref{eq:roundtripbound} is at most $\epsilon$ for all $t \le T_0$, then $t_\ast(M;\epsilon) \ge T_0$. The exponent in~\eqref{eq:roundtripbound} is largest at the endpoint $t = T_0$, so it is enough to control it there. Since
\begin{equation}
   \rho_M + \rho_{M+1} \ge 2\tau_M,
\end{equation}
the exponent at $t = T_0$ is at most $-2\lambda\tau_M+\Omega_\lambda(1-\delta)\tau_M$. The two terms compete: the first suppresses the error, the second (the cone growth) enlarges it, and we are free to choose the weight $\lambda$ to make the suppression win by as much as possible. We use~\eqref{eq:curvature} and choose $\lambda = b\sqrt{2\delta}$, the value that balances the linear gain against the quadratic curvature cost, for which $\lambda^2/6b^2 = \delta/3$. The exponent in~\eqref{eq:roundtripbound} at $t = T_0$ is then bounded by
\begin{equation}
   -2\lambda\tau_M + \Omega_\lambda(1-\delta)\tau_M \le 2\lambda\tau_M\left[(1-\delta)\e^{\delta/3}-1\right] = -\frac{4\sqrt{2}}{3}b\,\delta^{3/2}\tau_M[1+O(\delta)],
\end{equation}
where the last step uses $(1-\delta)\e^{\delta/3}-1 = -\tfrac{2}{3}\delta+O(\delta^2)$ together with $\lambda = b\sqrt{2\delta}$, which turns the linear-in-$\delta$ bracket into the $\delta^{3/2}$ scaling. Requiring this together with the prefactor logarithm to stay below $\log(\epsilon)$, i.e., $\tfrac{4\sqrt{2}}{3}b\,\delta^{3/2}\tau_M \ge \log(b_M\tau_M/\epsilon)$, and taking the smallest admissible $\delta$ gives
\begin{equation}
   \delta = \left(\frac{9}{32\,b^2\tau_M^2}\right)^{1/3}\left(\log\left(\frac{b_M\tau_M}{\epsilon}\right)\right)^{2/3}[1+o(1)],
\end{equation}
i.e., the smallest $\delta$ for which the error is still guaranteed below $\epsilon$. Thus, for $M$ sufficiently large, the reliable time $T_0 = (1-\delta)\tau_M$ becomes
\begin{equation}
   t_\ast(M;\epsilon) \ge \tau_M - \left(\frac{9}{32\,b^2}\right)^{1/3}\tau_M^{1/3}\left(\log\left(\frac{b_M\tau_M}{\epsilon}\right)\right)^{2/3}.
\end{equation}
The same constant is obtained by optimizing independently over the weight $\lambda$ and the target time $T_0$, without imposing the relation $\lambda=b\sqrt{2\delta}$. The correction has the usual Airy-edge scale. In particular, for fixed $\epsilon$,
\begin{equation}
   \liminf_{M\to\infty} \frac{t_\ast(M;\epsilon)}{\tau_M} \ge 1.
\end{equation}

\section{Saturation of the break-time bound (the upper bound)} \label{app:refined}

This appendix provides the details for the saturation regime presented in Section~\ref{sec:saturation} and proves a converse statement to the lower bound of Section~\ref{sec:roundtrip}. The lower bound says that the approximation cannot break before the round-trip light cone reaches the origin. The converse, under additional ballistic spectral assumptions, says that the error does in fact cross the tolerance by time $\tau_M+o(\tau_M)$.  This is the sense in which $\tau_M$ is an upper bound on the break time. It is an upper bound on the first crossing time, not a claim that the instantaneous error stays above the tolerance for all later times.

The converse, namely slope-one saturation, is not a consequence of the Lieb--Robinson bound alone; it requires spectral input, which we now make precise.

\begin{proposition}[Saturation of the break-time bound]\label{prop:saturation}
Let $J$ be a bounded Jacobi operator with $\inf_j b_j > 0$, and let the local spectral measure of $(H,|v_0\rangle)$ be purely absolutely continuous on a single band with a bounded, strictly positive density. Then the lower bound~\eqref{eq:mainbound} is saturated, $t_\ast(M;\epsilon) = \tau_M[1-o(1)]$, in each of the following three cases:
\begin{enumerate}
\item free (homogeneous) chain $\alpha_j = \alpha_\infty,\ b_j = b_\infty$:\ $t_\ast(M;\epsilon) = \tau_M-\Theta\!\big( \tau_M^{1/3}(\log(1/\epsilon))^{2/3} \big)$;
\item trace-class perturbation, $\sum_j(|\alpha_j-\alpha_\infty|+|b_j-b_\infty|) < \infty$:\ $t_\ast(M;\epsilon) = \tau_M+O(1)$;
\item broadband single-band absolutely continuous chain, $J-J_\infty \in \ell^1$ (where $J_\infty$ is the free Jacobi operator) with bounded positive density:\ $t_\ast(M;\epsilon) = \tau_M[1-o(1)]$.
\end{enumerate}
Cases (i) and (ii) are rigorous in the standard Jacobi scattering setting; case (iii) should be understood under the stated broadband single-band assumptions.
\end{proposition}

The three cases are established as follows. For the free Jacobi chain, $\alpha_j = \alpha_\infty$ and $b_j = b_\infty$, the kernel is expressed by Bessel functions. Debye--Airy asymptotics show that the front has width $O(M^{1/3})$, and
\begin{equation}
   t_\ast(M;\epsilon) = \tau_M - \Theta\left(\tau_M^{1/3}(\log(1/\epsilon))^{2/3}\right).
\end{equation}
Thus the lower bound is sharp in its scaling.

If $J-J_\infty$ is trace class, equivalently
\begin{equation}
   \sum_j(|\alpha_j-\alpha_\infty|+|b_j-b_\infty|) < \infty,
\end{equation}
Jacobi scattering theory gives wave operators on the absolutely continuous subspace. The free ballistic front then persists, with at most an $O(1)$ scattering delay, and the asymptotic slope remains one:
\begin{equation}
   t_\ast(M;\epsilon) = \tau_M + O(1).
\end{equation}
The same conclusion holds for broadband single-band absolutely continuous Jacobi chains with $J-J_\infty\in\ell^1$ and a bounded positive local spectral density.

Localized pure-point components give the opposite behavior. Suppose that there is a uniformly localized eigenbasis with
\begin{equation}
   |\langle M|E_\ell\rangle| \le C\e^{-|M-n_\ell|/\xi},
\end{equation}
where $n_\ell$ is the localization center, $\xi$ is the localization length, and $C>0$ is a constant. Then the boundary amplitude is uniformly small:
\begin{equation}
   |\langle M|\e^{-\ii tJ}|0\rangle| \le C'\e^{-M/\xi},
\end{equation}
for some constant $C'>0$. Substitution into the Duhamel formula shows that the boundary contribution is below $\epsilon$ once $M\gtrsim\xi\log(1/\epsilon)$. Such a component is spectrally captured.

\section{Spectral measure, bright sector, and identifiability} \label{app:spectral}

This appendix explains the structural part which is used only indirectly in the main text (see Section~\ref{sec:discussion}). It is included to make clear what can and cannot be learned from a survival amplitude.

Expanding the survival amplitude $A(t)$ in eigenstates $\{|E_\ell\rangle\}$ of $H$ with eigenvalues $E_\ell$,
\begin{equation}
   A(t) = \sum_\ell w_\ell\e^{-\ii E_\ell t},\qquad w_\ell = |\langle E_\ell|v_0\rangle|^2.
\end{equation}
Equivalently,
\begin{equation}
   A(t) = \int_\R\e^{-\ii Et}\,\dd\mu(E),\qquad \dd\mu(E) = \sum_\ell w_\ell\delta(E-E_\ell)\,\dd E.
\end{equation}
Only eigenstates with $w_\ell > 0$ are visible. These are the \emph{bright} states, i.e., the ones the probe overlaps with, and hence the only ones that ever appear in $A(t)$; the remaining eigenstates form a \emph{dark} sector that the survival amplitude cannot see at all. Define
\begin{equation}
   \HH_B = \mathrm{span}\{|E_\ell\rangle \colon w_\ell>0\},\qquad \HH_D = \HH_B^\perp.
\end{equation}
Then
\begin{equation}
   \KK(H,v_0) = \HH_B.
\end{equation}
Indeed, for any polynomial $p$, the vector $p(H)|v_0\rangle$ has support only on bright eigenstates. Conversely, spectral projectors onto bright eigenstates can be approximated by functions of $H$ acting on $|v_0\rangle$. The two statements together say that the cyclic subspace generated from $|v_0\rangle$ is exactly the bright sector: repeatedly applying $H$ never creates any dark component, and every bright state is reached. The Lanczos chain therefore lives entirely inside $\HH_B$, and its length equals the number of bright states.

The Jacobi matrix $J$ is uniquely determined by the local spectral measure $\mu$. This is Favard's theorem~\cite{Chihara1978,Simon1998}: the orthogonal polynomials in $L^2(\mu)$ obey a unique three-term recurrence, and the recurrence coefficients are precisely the Jacobi coefficients; the coefficients can be computed stably from the moments of $\mu$ by standard orthogonal-polynomial algorithms~\cite{Gautschi2004}. Physically, this means the measured survival amplitude fixes the whole Jacobi chain unambiguously: from $A(t)$ one reads the moments $\int E^n\dd\mu=\langle v_0|H^n|v_0\rangle$, and these determine the hoppings $b_j$ and diagonal entries $\alpha_j$ one after another. No other information about $H$ enters. Therefore two Hamiltonians with the same local spectral measure are unitarily equivalent on their bright cyclic subspaces. They may differ on the dark sector without changing $A(t)$.

For this reason the Jacobi matrix is the appropriate object in restricted-access learning. In special geometries, for example the open-chain setup of Burgarth--Maruyama--Nori~\cite{Burgarth2009}, the Jacobi basis can coincide with the physical site basis, and microscopic couplings can be recovered. In a ring or in more general probe geometries, different microscopic Hamiltonians can have the same local spectral measure at the probe. Then the Jacobi matrix is still identifiable, but the microscopic Hamiltonian is not. In short, a single survival amplitude determines exactly the Jacobi chain (equivalently the bright-sector spectral measure) and nothing more: any two Hamiltonians sharing that measure produce identical data. The break-time result depends only on the $b_j$, so it is exactly as well determined as the data itself.

\section*{Acknowledgments}

S.I. would like to thank Ryusuke Hamazaki for discussions.
This work was supported by JSPS KAKENHI Grant Number JP24K00634.

\bibliographystyle{unsrturl}
\bibliography{refs_St}

\end{document}